\documentclass[twocolumn,aps,prapplied]{revtex4-2}
\usepackage{amsmath}
\usepackage{amssymb}
\usepackage{graphicx}

\makeatletter

\begin{document}
\title{Origin of the repulsive Casimir force in giant polarization-interconversion materials}

\author{Zhou Li$^{1,2,3}$, Chinmay Khandekar$^{1}$}
\affiliation{$^{1}$ Purdue University, West Lafayette, Indiana 47906, United States}
\affiliation{$^{2}$ GBA Branch of Aerospace Information Research Institute, Chinese Academy of Sciences, Guangzhou 510535, China}
\affiliation{$^{3}$ University of Chinese Academy of Sciences, Beijing 100039, China}
\begin{abstract}
Achieving strong repulsive Casimir forces through engineered coatings can pave the way  for micro- and nano-electromechanical applications where adhesive forces currently cause reliability issues. Here, we exploit Lifshitz theory to identify the requirements for repulsive Casimir forces in gyrotropic media for two limiting cases (ultra-strong gyroelectric and non-gyroelectric).  We show that the origin of repulsive force in media with strong gyrotropy such as Weyl semi-metals arises from the giant interconversion of polarization of vacuum fluctuations.
\end{abstract}
\date{\today}
\maketitle

\section{Introduction}
The Casimir force \cite{Cas,Lifshitz,Lamo1,Klim,Lamo2} exists between charge-netural bodies separated by submicron gaps because of the quantum fluctuations of electromagnetic fields. It competes with other forces in the micro-to-nano-meter region. For majority of geometric and material configurations, the Casimir force is known to be attractive. Realizing repulsive Casimir force is not only fundamentally important but also technologically relevant for micro- and nano-electromechanical systems (MEMS and NEMS). From an engineering perspective, attractive Casimir force dominates in the sub-micrometer regime where closely spaced system parts tend to attract each other. Our findings of the repulsive Casimir force could be applied in the design of MEMS and NEMS, to avoid attraction-induced friction. Our findings of the mixed attractive-repulsive Casimir force could be applied to extract energy from vacuum fluctuations.

The most-studied geometry in the context of Casimir force \cite{Nori1,Nori2,Nori3} is that of two parallel plates of real, dispersive materials separated by a vacuum gap, where the force is accurately described by Lifshitz theory based on the fluctuation--dissipation theorem \cite{Lifshitz,Klim}. One well-known approach to obtain repulsive Casimir force \cite{Milton} is to use two plates of dielectric constants $\epsilon_1, \epsilon_2$ separated by a liquid medium of permittivity $\epsilon_l$ instead of vacuum such that $\epsilon_{1}<\epsilon_{l}<\epsilon_{2}$. Recent works have revealed other approaches such as the use of Teflon-coated metallic plates~\cite{Zhang}. Another intriguing approach relies on exploiting topological materials to achieve repulsive Casimir forces using topological materials. These include three dimensional topological insulators and Weyl semimetals \cite{Grushin,Wilson,Jiang,Zyuzin}, two dimensional Chern insulators and the Graphene family \cite{Tse,Pablo,Pablo2}. However, the underlying mechanism which gives rise to repulsive Casimir forces in these topological materials has remained unexplored.

In this paper, we elucidate the origin of Casimir repulsion between two ultra-strong gyrotropic plates (Eq.~(3)).  Weyl semimetals are strong (not ultra-strong) gyrotropic medium, the Casimir force has to be determined numerically. For the repulsive Casimir force, we choose the typical parameters of the dielectric tensor of a Weyl semimetal, which could be realized in machine-learning assisted material growth.
\begin{figure}[t!]
\centering\includegraphics[width=\linewidth]{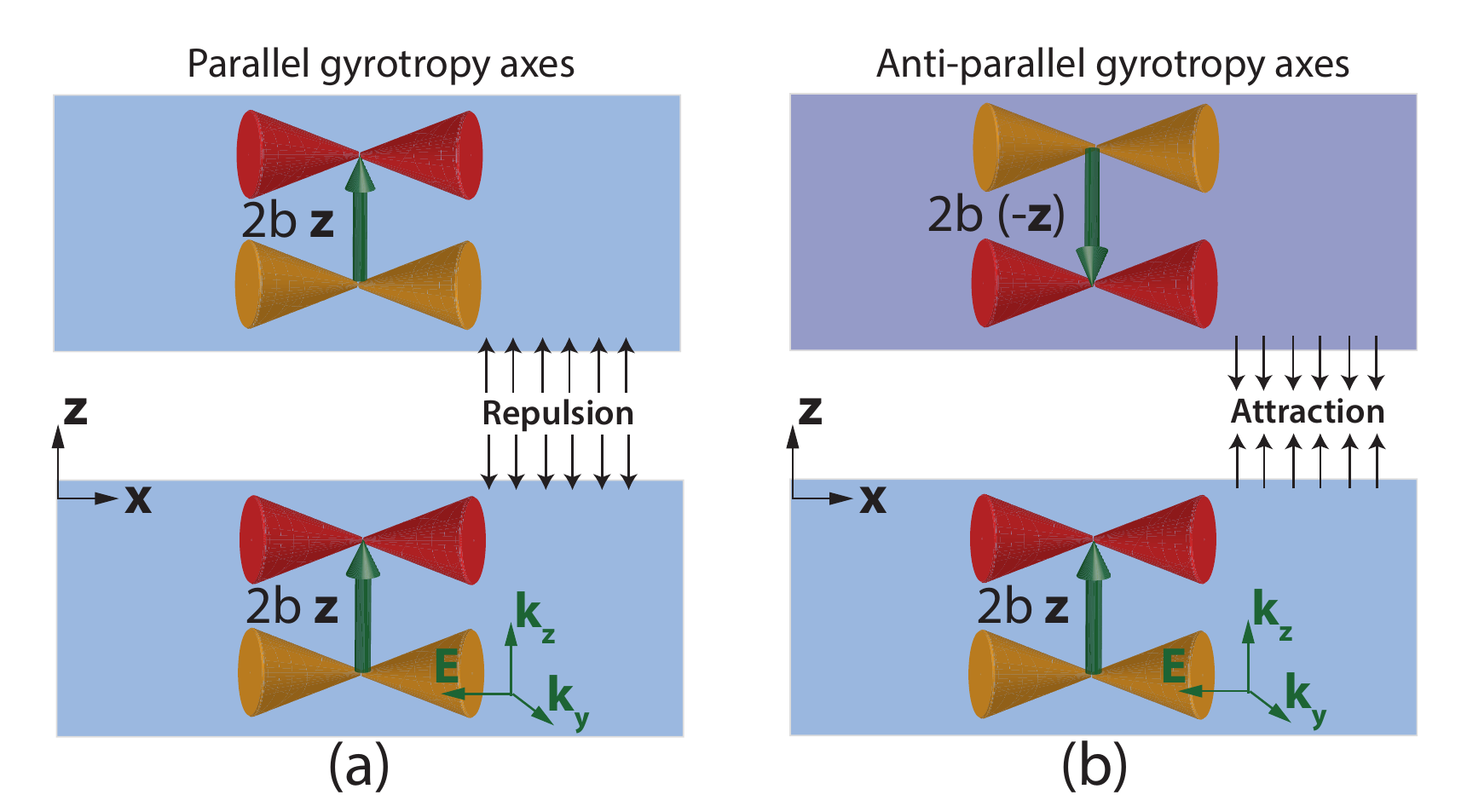}
\caption{Weyl semimetals are strong gyrotropic media whose gyrotropy axis is along the direction of the associated momentum-separation vector $b$ of two Weyl nodes. (a) We show that for two plates with parallel gyrotropy axes, the Casimir force between the plates can become repulsive. (b) For plates with antiparallel gyrotropy axes, the Casimir force is always attractive. 
}\label{scm1}
\end{figure}
Strong gyrotropic media have recently opened up new promising fundamental and technological avenues for thermal radiation-based devices\cite{Fan1,Fan2,Chinmay} as well. 

\section{Formalism}
We start from the Lifshitz theory of the Casimir free energy for two parrallel plates separated by a vacuum gap, which is 
\begin{equation}
\frac{E_{c}(d)}{A}=k_{B}T\times\sum_{n}\int\frac{d^{2}k}{(2\pi)^{2}}\mathrm{Ln}[\mathrm{det}(1-\mathbf{R_{1}R_{2}}e^{-2k_{1}d})]
\end{equation}
where $R_{1}$ and $R_{2}$ are the reflection matrix for the plate
1 and plate 2, $E_{c}(d)$ is the free energy, $A$ is the area, $k_{1}$
is the absolute value of the imaginary part of the wave-vector $k_{z}$,
$T$ is the temperature and $k_{B}$ is the Boltzman constant. The
sum over n is defined as $1/2(n=0)+\sum_{n>0}$ with n the index of the Matsubara frequencies $\xi_{n}=2\pi nk_{B}T/\hbar$.
A reflection matrix for an electromagnetic wave injecting from one
medium (e.g. the vacuum) to another medium (e.g. a silicon plate or
gyrotropic plate) is defined as, 
\[
\mathbf{R}=\left[\begin{array}{cc}
r_{ss} & r_{sp}\\
r_{ps} & r_{pp}
\end{array}\right]
\]
where $r_{ss}=R_{TE}/A_{TE}$, $r_{sp}=R_{TM}/A_{TE}$, $r_{pp}=R_{TM}/A_{TM}$
and $r_{ps}=R_{TE}/A_{TM}$. Here $A_{TM}$($R_{TM}$) are the amplitudes
of injecting (reflected) transverse magnetic (TM) mode, and $A_{TE}$($R_{TE}$)
are the amplitudes of injecting (reflected) transverse electric (TE)
mode. While the TE mode and the TM mode are orthogonal electromagnetic
wave-functions traveling in a vacuum, in other media, the electromagnetic
eigenstates may be a mixture of TE and TM modes. Thus at the boundary
of a vacuum and another medium, the TE mode to TM mode transfer ratio
($r_{ps}$ and $r_{sp}$) could be non-zero. The Casimir force is
the derivative of the Casimir energy, $F=-\frac{\partial E_{c}(d)}{\partial d}$.
If $F>0$, the force is repulsive, if $F<0$, the force is attractive.
The Casimir pressure is defined as 
\begin{equation}
P=F/A=k_{B}T\times\sum_{n}\int\frac{d^{2}k}{(2\pi)^{2}}\left[-\frac{\partial \mathrm{Ln}(L)}{\partial d}\right]
\end{equation}
where $L=\mathrm{det}[1-\mathbf{R_{1}R_{2}}e^{-2k_{1}d}]$. While
in general the repulsive Casimir force should be verified numerically,
in two limit cases (ultra-strong-gyrotropic and non-gyrotropic), we provide here the requirement for a repulsive
Casimir force in its basic form, with details given in the appendix.
For a ultra-strong gyrotropic material, the reflection
coefficients $|r_{sp}|$ and $|r_{ps}|$ are much larger than
$|r_{ss}|$ and $|r_{pp}|$, the requirement for a repulsive force
is 
\begin{equation}
\sum_{n}\int\frac{d^{2}k}{(2\pi)^{2}}(r_{sp1}r_{ps2}+r_{ps1}r_{sp2})<0
\label{conditionrepulsive}
\end{equation}
For a non-gyrotropic isotropic medium, $r_{sp1}=r_{ps1}=0$ and $r_{sp2}=r_{ps2}=0$,
the requirement for a repulsive force is 
\begin{equation}
\sum_{n}\int\frac{d^{2}k}{(2\pi)^{2}}(r_{ss1}r_{ss2}+r_{pp1}r_{pp2})<0
\end{equation}
This requirement also holds for the weak gyrotropic material where
$|r_{sp}|$ and $|r_{ps}|$ are much smaller than $|r_{ss}|$ and
$|r_{pp}|$.

Based on the Dirac-Maxwell correspondence, we define a Maxwell Hamiltonian and obtain the reflection coefficients ($r_{ss}$,$r_{pp}$,$r_{sp}$,$r_{ps}$) by connecting the wave-functions (eigenstates) at the interface. The Maxwell equations for a specific medium (omitting the magneto-electric media) is
given by 
\begin{equation}
\left[\mathbf{\begin{array}{cc}
\epsilon & 0\\
0 & \mu
\end{array}}\right]\frac{\partial}{\partial t}\left[\begin{array}{c}
\mathbf{E}\\
\mathbf{H}
\end{array}\right]=\left[\begin{array}{c}
\nabla\times\mathbf{H}\\
-\nabla\times\mathbf{E}
\end{array}\right]
\end{equation}
where $\mathbf{E}$ and $\mathbf{H}$ are the electric and magnetic
fields. Note that $(\mathbf{S}\cdot\nabla)\mathbf{H}=i\nabla\times\mathbf{H}$
where $\mathbf{S}$ is the spin-1 matrices with $S_{x}$, $S_{y}$
and $S_{z}$ defined as, 
\[
S_{x}=\left[\begin{array}{ccc}
0 & 0 & 0\\
0 & 0 & -i\\
0 & i & 0
\end{array}\right],S_{y}=\left[\begin{array}{ccc}
0 & 0 & i\\
0 & 0 & 0\\
-i & 0 & 0
\end{array}\right],S_{z}=\left[\begin{array}{ccc}
0 & -i & 0\\
i & 0 & 0\\
0 & 0 & 0
\end{array}\right]
\]
Maxwell equations transform into a Dirac-like equation \cite{Nori,Zubin,Zubin1}.
Assuming a plane wave $e^{i\mathbf{k\cdot r}-i\omega t}$ of the electromagnetic field, we obtain the Maxwell Hamiltonian ( $\omega=i\xi_{n}$)
\begin{equation}
H_{Max}=\left[\mathbf{\begin{array}{cc}
0 & \mathbf{\epsilon^{-1}\mathbf{S\cdot k}}\\
-\mu^{-1}\mathbf{S\cdot k} & 0
\end{array}}\right]\label{Max}
\end{equation}
For the Casimir force, the permittivity matrix at imaginary Matsubara frequencies $\omega=i\xi_{n}$ is relevant. With gyrotropy axis along the z-direction, the permittivity matrix and its inverse are given below:
\begin{equation}
\epsilon=\left[\begin{array}{ccc}
\epsilon_{1} & g & 0\\
-g & \epsilon_{1} & 0\\
0 & 0 & \epsilon_{2}
\end{array}\right],\epsilon^{-1}=\left[\begin{array}{ccc}
d_{1} & g' & 0\\
-g' & d_{1} & 0\\
0 & 0 & d_{2}
\end{array}\right]
\label{gyroperm}
\end{equation}
Both are real-valued at imaginary Matsubara frequencies. Here
$d_{1}=\epsilon_{1}/(\epsilon_{1}^{2}+g^{2})$, $d_{2}=1/\epsilon_{2}$
and $g'=-g/(\epsilon_{1}^{2}+g^{2})$. For a plane wave travelling
in the $k_{x}-k_{z}$ plane, with the wave vector $\mathbf{k}=(k_{x},0,q)$,
the eigenvalue is given by  
\begin{equation}
\xi^2_{n}=-\left[(d_{1}+d_{2})k_{x}^{2}+2d_{1}q^{2}\mp\sqrt{P}\right]/2
\end{equation}
here  $P=k_{x}^{4}(d_{1}-d_{2})^{2}-4(k_{x}^{2}+q^{2})q^{2}g'^{2}$ and we set $c=1$, so $\xi_{n}/c$ is written as $\xi_{n}$. We obtain
an inverse solution of $q$ from $\xi_{n}$ and $k_{x}$,which is
\begin{eqnarray}
q^{2}&=&1/[2(d_{1}^{2}+g'^{2})]\times \notag \\
&&\left[-d_{1}[(d_{1}+d_{2})k_{x}^{2}+2\xi_{n}^{2}]-k_{x}^{2}g'^{2}\pm\sqrt{\triangle}\right]\label{disp}
\end{eqnarray}
where 
\begin{equation}
\triangle=[d_{1}(d_{1}-d_{2})+g'^{2}]^{2}k_{x}^{4}-4g'^{2}\xi_{n}^{2}(\xi_{n}^{2}+d_{2}k_{x}^{2})
\end{equation}
This gives four solutions (eigenstates) of electromagnetic waves inside
a gyroelectric medium. However, along one specific direction, only
two solutions are allowed. The two momenta $q_{1}$ and $q_{2}$ are
chosen by the following rules: along the direction $+z$, the plane
waves should decay at infinity. This requires the imaginary part $\mathrm{Im}(q)>0$,
but the real part is not restricted, $\mathrm{Re}(q)>0$ or $\mathrm{Re}(q)<0$ ; along
the direction $-z$, it requires the imaginary part $\mathrm{Im}(q)<0$, with
no restriction on the real part. In both cases we have $q_{1}=-q_{2}^{*}$.
The eigenstate is a function of the momenta $q_{1}$
and $q_{2}$, $\psi_{1}=[e_{x1},e_{y1},e_{z1},h_{x1},h_{y1},h_{z1}]^{T}=[E_{1},H_{1}]^{T}$
and $\psi_{2}=[e_{x2},e_{y2},e_{z2},h_{x2},h_{y2},h_{z2}]^{T}=[E_{2},H_{2}]^{T}$,
given by $E_{1,2}=E(q_{1,2})$ and $H_{1,2}=H(q_{1,2})$, with 
\begin{equation}
E(q)=\left[E_x,-g'q\xi^2_{n},-d_{2}k_{x}(d_{1}k_{x}^{2}+d_{1}q^{2}+\xi_{n}^{2})\right]
\end{equation}
where $E_x=q[(d_{1}^{2}+g'^{2})(k_{x}^{2}+q^{2})+d_{1}\xi_{n}^{2}]$ and 
\begin{equation}
H(q)=i\xi_{n}\left[-g'q^2,d_{1}k_{x}^{2}+d_{1}q^{2}+\xi_{n}^{2},g'k_{x}q\right]
\end{equation}
The details of obtaining the reflection coefficients are given
in the appendix. For a non-gyroeletric medium, $g'=0$ and $d_{1}=d_{2}$,
we obtain $q=iq_{I}$ from Eq.~(\ref{disp}) where $q_{I}^{2}=k_{x}^{2}+\xi_{n}^{2}/d_{1}=\varepsilon(i\xi_{n})\xi_{n}^{2}/c^{2}+k_{x}^{2}$.

\section{Gyroelectric medium and the magneto-plasma model}

\begin{figure}[tp]
\begin{centering}
\includegraphics[width=3.5in,height=3in] {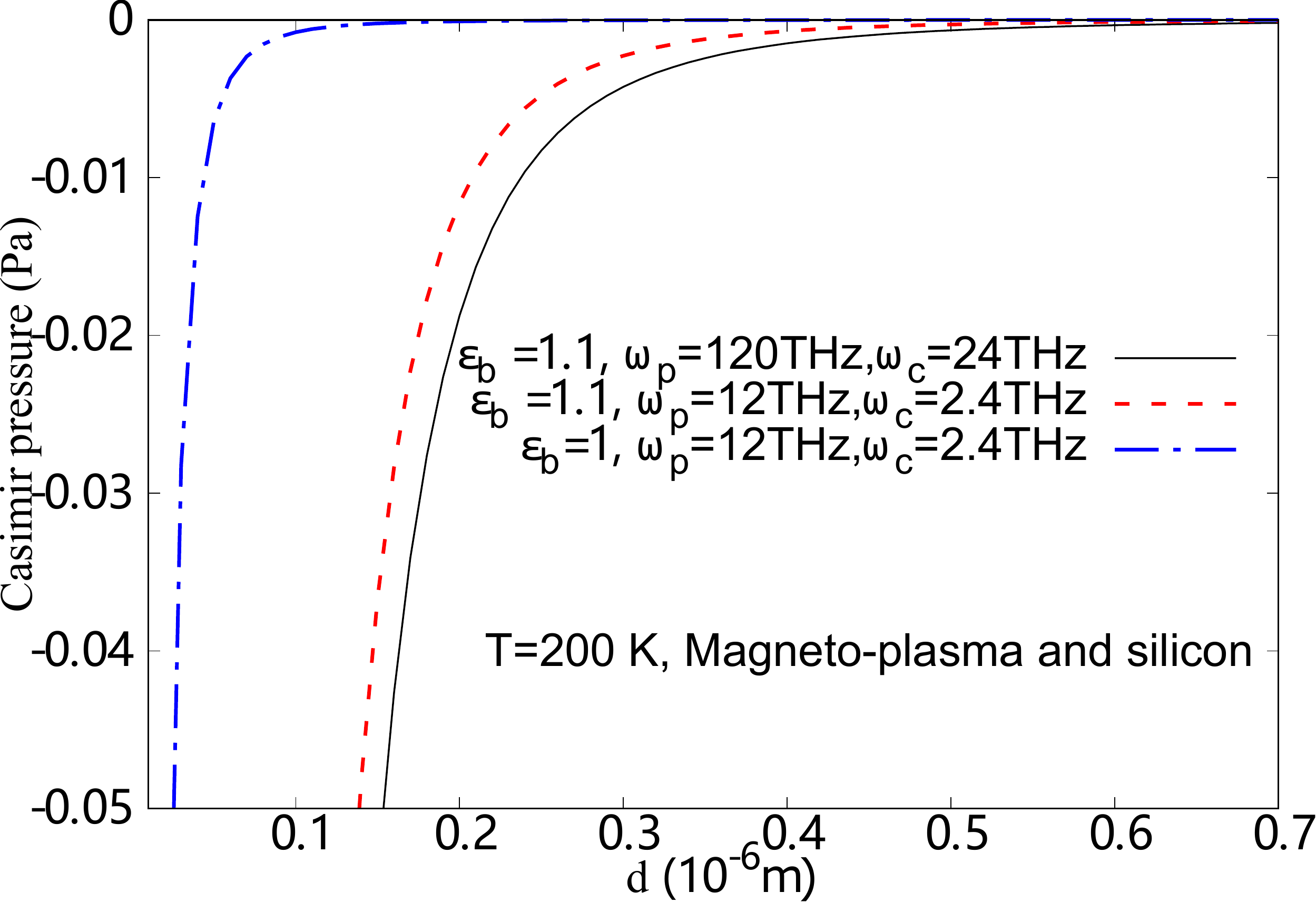} 
\par\end{centering}
\caption{(Color online) Casimir pressure as a function of the distance, $d$, between a gyrotropic plate and a silicon
plate. The dielectric constant of the gyrotropic plate is determined from the magneto-plasma model. Note that the Casimir force is always negative (attractive). When $\epsilon_{b}$ decreases from 1.1 to 1, the Casimir
force is greatly suppressed.}
\label{fig2} 
\end{figure}

\begin{figure}[tp]
\begin{centering}
\includegraphics[width=3.5in,height=3in]{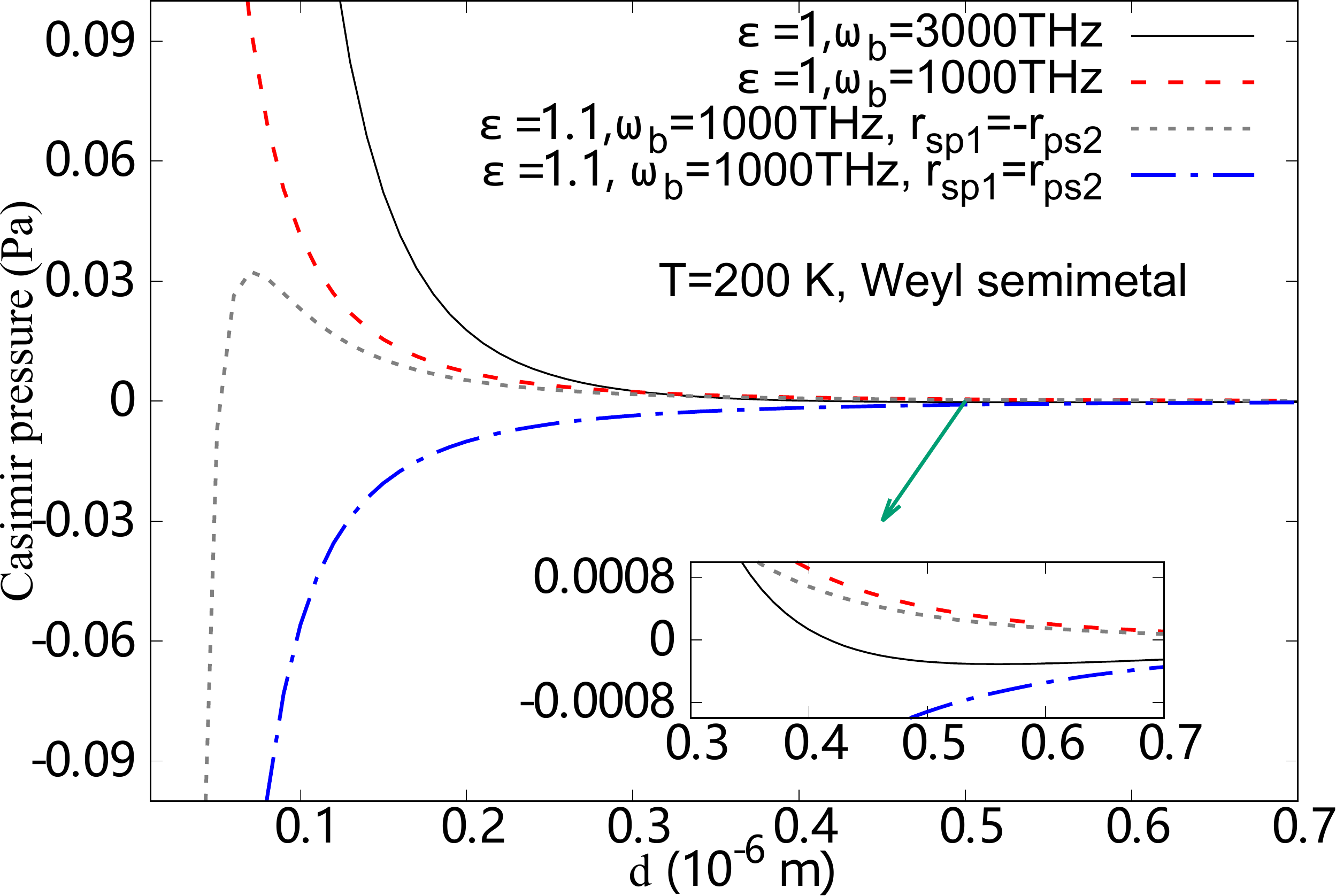} 
\end{centering}
\caption{(Color online) Casimir pressure as a function of the distance, $d$,
between two Weyl-semimetal plates. Note that for the blue dash dotted curve, the two gyro-axis are anti-parallel, we have $r_{sp1}=r_{ps2}$, and the Casimir force is always attractive. For the other three curves, the two gyro-axis are parallel, we have $r_{sp1}=-r_{ps2}$, the Casimir force is tuned from attractive to repulsive. The grey dotted curve crosses zero at around $d=0.052\mu m$ (unstable equilibrium) and the black solid curve crosses zero at around $d=0.41\mu m$ (stable equilibrium). For the red dashed curve, the Casimir force is always repulsive.}
\label{fig2} 
\end{figure}

Now we consider a gyroelectric medium, which has non-zero off-diagonal elements in a permittivity matrix,
typically given by the following magneto-plasma model: 
\begin{equation}
\epsilon=\epsilon_{b}-\frac{\omega_{p}^{2}}{(\omega+i\Gamma)^{2}-\omega_{c}^{2}}\left[\begin{array}{ccc}
1+i\frac{\Gamma}{\omega} & -i\frac{\omega_{c}}{\omega} & 0\\
i\frac{\omega_{c}}{\omega} & 1+i\frac{\Gamma}{\omega} & 0\\
0 & 0 & \frac{(\omega+i\Gamma)^{2}-\omega_{c}^{2}}{\omega(\omega+i\Gamma)}
\end{array}\right]
\end{equation}
Here $\omega_{p}$ is the plasma frequency, $\omega_{c}$
is the cyclotron frequency.
$\Gamma$ is the inverse lifetime of the charge carriers inside the
medium, microscopically determined by the scattering process from
impurities, phonons or other sources. The background dielectric constant
$\epsilon_{b}=\epsilon_{\infty}+\epsilon_{inter}+\epsilon_{lattice}$,
where $\epsilon_{\infty}$ is the high-frequency limit, $\epsilon_{inter}$ and $\epsilon_{lattice}$
are contributions from inter-band transitions and lattice vibrations, respectively \cite{Fan1}. In the Matsubara frequency domain
$\omega=i\xi_{n}$, the permittivity matrix (Eq.\ref{gyroperm}) is described by diagonal coefficients $\epsilon_{1}=\epsilon_{b}+\frac{\omega_{p}^{2}(1+\Gamma/\xi_{n})}{(\xi_{n}+\Gamma)^{2}+\omega_{c}^{2}}$,
$\epsilon_{2}=\epsilon_{b}+\frac{\omega_{p}^{2}}{\xi_{n}(\xi_{n}+\Gamma)}$
and the off-diagonal coefficient $g=\frac{-\omega_{p}^{2}(\omega_{c}/\xi_{n})}{(\xi_{n}+\Gamma)^{2}+\omega_{c}^{2}}$.

\section{Strong gyroelectric medium and the Weyl semimetal}

The constitutive relations for an ideal Weyl semimetal is given by
\cite{Kotov,Soh}
\begin{equation}
D =  \epsilon_{w}\mathbf{E}+\frac{ie^{2}}{4\pi^{2}\hbar\omega}(2\mathbf{b}\times\mathbf{E}-2b_{0}\mathbf{B)}
\end{equation}
where $ \epsilon_{w}$ ($\epsilon$ in Fig.~3 and Fig.~4) is the diagonal part of the dielectric constant of the Weyl semimetal. The first term in the bracket comes from the anamolous Hall effect, and contributes to the off-diagonal part of the dielectric constant. The second term in the bracket comes from the chiral magnetic effect, for simplicity we set $b_0=0$. Here $\mathbf{b}$ is the momentum-separation of the
two Weyl nodes, and we choose $\mathbf{b}=bk_{z}$ to be along the
z-direction. The typical frequency $\omega_{b}=\frac{e^{2}b}{2\pi^{2}\hbar\epsilon_{0}}$
is defined from the anamolous Hall effect, where $\epsilon_{0}$ is
the vacuum permittivity. For $b=0.5\mathring{A}^{-1}$, we have $\omega_{b}=6949$
THz. The microscopic theory of obtaining these parameters is provided
in \cite{Burkov,Zyuzin2,Zhou1,Zhou2,Ashby1,Ashby2,Rodi1,Rodi2}. For the longitudinal
optical conductivity, one finds that the optical conductivity increases
linearly as a function of the frequency $\omega$ \cite{Ashby1}. Similarly
it increases sub-linearly in topological insulators \cite{Zhou2}. After dividing by $\omega$, the dielectric constant $ \epsilon_{w}$ should not change too much as a function of $\omega$, therefore, in the Matsubara frequency domain, we choose $\epsilon_{1}=\epsilon_{2}\simeq1$(see discussions below equation (11) in \cite{Zyuzin}), and the off-diagonal
part of the permittivity matrix to be $g=\frac{\omega_{b}}{\xi_{n}}$. For $n=0$, we add a tiny positive number to $\xi_{0}$ to avoid the divergence.

\begin{table}
\caption{\label{Parameter} Off-diagonal permittivity ($g$) in comparison with diagonal permittivity $\epsilon_1$ for magneto-plasma model and Weyl semimetal.}
\begin{tabular}{c|c|c}
\hline 
\hline
 & $g (n=1)$  & $g (n=2)$ \\
\hline 
Magneto-plasma model & 0.076 $\ll \epsilon_1$   & 0.0096 $ \ll \epsilon_1$ \\
\hline 
Weyl semimetal & 6.0778  $\gtrsim \epsilon_1$ & 3.039 $\gtrsim \epsilon_1$ \\
\hline 
\end{tabular}
\end{table}
In Table~I we show the absolute values of the gyrotropic coupling (off-diagonal permittivity) for the Matsubara frequencies corresponding to $n=1,2$ with the diagonal permittivity $\epsilon_1 \approx 1$ for both models. For the magneto-plasma model we use $\omega_{p}=120$ THz and $\omega_{c}=24$ THz and for the Weyl semimetal  we use $\omega_{b}=1000$ THz. The temperature is assumed to be T=200K.

\section{Numerical results}

\begin{widetext}

\begin{figure}[tp]
\begin{centering}
\includegraphics[width=6in,height=5in,keepaspectratio]{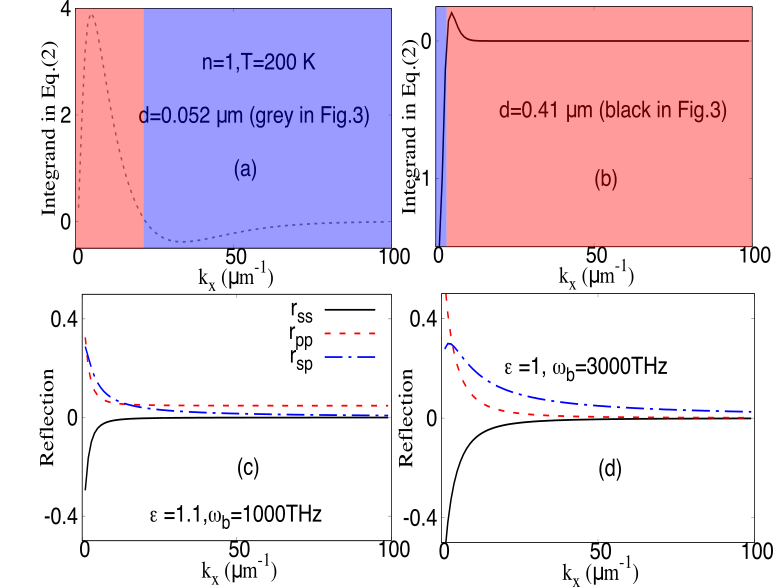} 
\par\end{centering}
\caption{(Color online) The integrand ($k_x\times[-\frac{\partial \mathrm{Ln}(L)}{\partial d}]$) in Eq.~(2) (shown in a,b) and the associated reflection coefficients (shown in c,d)
as a function of $k_x$. We set the Matsubara frequency at $n=1$ and choose parameters the same as those of the black and grey curves in Fig.~3. We use the red (blue) color to denote the repulsive (attractive) contribution to the Casimir force respectively in (a) and (b). }
\label{fig3} 
\end{figure}
\end{widetext}

In Fig.~2, we show that gyrotropy i.e non-reciprocity is not a sufficient condition for repulsive Casimir force. We present numerical results of the Casimir force (pressure), between a silicon plate
and a gyrotropic (magneto-plasma) plate which is clearly attractive. The frequency-dependent dielectric constant
of silicon is given by $\epsilon_{Si}=\epsilon_{(Si)\infty}+(\epsilon_{(Si)0}-\epsilon_{(Si)\infty})\frac{\omega_{0}^{2}}{\xi_{n}{}^{2}+\omega_{0}^{2}}+\frac{\omega_{p}^{2}}{\xi_{n}(\xi_{n}+\Gamma)}$,
where $\omega_{0}=6.6\times10^{15}$Hz, $\omega_{p}=3.6151\times10^{14}$Hz,
$\Gamma=7.868\times10^{13}$Hz, $\epsilon_{(Si)\infty}=1.035$, and $\epsilon_{(Si)0}=11.87$.
As $\epsilon_{b}$ moves toward 1, the attractive Casimir force is greatly suppressed. In comparison, the Casimir pressure between two perfect conducting plates is $P_C=-\frac{\hbar c \pi^{2}}{240 d^4}$ as we derived in the appendix. At $d=0.2\mu m$, $P_C=-0.8$ Pa. 

In Fig.~3 we present numerical results of the Casimir force (pressure), between two Weyl-semimetal plates. Because of strong gyrotropy as evident from Table~I, vacuum fluctuations in the gap between the plates experience strong polarization conversion. The associated reflection coefficients satisfy the condition given by Eq.~\ref{conditionrepulsive} for a suitable range of in-plane wavevectors leading to repulsive Casimir force between the plates. For the case of parallel gyrotropy axes, we plot the Casimir force with three different parameter sets represented by black, red and grey curves respectively. As the distance $d$ is changed, the Casimir force is tuned from repulsive to attractive, or always repulsive, or tuned from attractive to repulsive, respectively. For the case of anti-parallel gyrotropy axes, the Casimir force is always attractive, as represented by the blue curve. For the black curve ($\epsilon=1$ and $\omega_{b}=3000$ THz), the Casimir force is tuned to be zero at $d_0=0.41\mu m$ (stable equilibrium); for the grey curve  ($\epsilon=1.1$ and $\omega_{b}=1000$ THz), the zero-Casimir-force point $d_0=0.052\mu m$ is an unstable equilibrium.

In Fig.~4 we investigate the origin of the stable and unstable equilibria, as shown in Fig.~3 by the black and grey curves respectively. As Casimir forces depend on vacuum fluctuations at all momenta and frequencies, their interpretation is significantly more challenging than conventional narrowband (eg: laser driven) coherent optical forces. Therefore, we separately identify the contribution of specific plane waves to explain the origin of the repulsive Casimir force. We present the integrand of Eq.~(2) in Fig.~4(b), as a function of $k_x$ and fix the Matsubara frequency at $n=1$. The black curve is negative (contribute to an attractive Casimir force) for small $k_x$ and positive (contribute to a repulsive Casimir force) for large $k_x$. The grey curve shows an opposite trend. In Fig.~4(c) and Fig.~4(d) we plot the amplitude of the reflection coefficients for the grey curve and black curve respectively. For Fig.~4(c), $|r_{sp}|$ is larger than $|r_{pp}|$ and $|r_{ss}|$ in a narrow range of $k_x$, 
while for Fig.~4(d) $|r_{sp}|$ is larger than $|r_{pp}|$ (or $|r_{ss}|$) in a broad range of $k_x$.

\section{Conclusion}
In this paper we develop a theory of Casimir force based on the Lifshitz formula and the Dirac-Maxwell correspondence. We choose a magneto-plasma model and a Weyl semimetal model to apply the theory. We find repulsive Casimir force in the latter, which is a strong gyrotropic medium. We also find the repulsive Casimir force is tuned from repulsive to zero at specific distance (equilibrium). This could be applied in the design of MEMS(NEMS) to reduce the friction between tiny parts of devices.

\section*{Acknowledgment}

The authors would like to acknowledge Zubin Jacob and Fanglin Bao for helpful discussions. Z. L. acknowledges the support of Chinese Academy of Science funding No. E1Z1D10200. This work is supported in part by the National Natural Science Foundation of China (No.61988102).

\appendix

\section{Details of the condition for a repulsive Casimir force from the reflection
matrix }
Start from the Lifthiz formula, the Casimir pressure is defined as 
\begin{equation}
P=F/A=k_{B}T\times\sum_{n}\int\frac{d^{2}k}{(2\pi)^{2}}\left[-\frac{\partial \mathrm{Ln}(L)}{\partial d}\right]
\end{equation}
where 
\begin{equation}
L=\mathrm{det}(1-\mathbf{R_{1}R_{2}}e^{-2k_{1}d})
\end{equation}
the product of the reflection matrices at interface 1 and 2 is given
by, 
\[
\mathbf{R_{1}R_{2}}=\left[\begin{array}{cc}
r_{ss1} & r_{sp1}\\
r_{ps1} & r_{pp1}
\end{array}\right]\left[\begin{array}{cc}
r_{ss2} & r_{sp2}\\
r_{ps2} & r_{pp2}
\end{array}\right]=\left[\begin{array}{cc}
D_{1} & D_{2}\\
D_{3} & D_{4}
\end{array}\right]
\]
where 
\begin{align*}
D_{1} & =r_{ss1}r_{ss2}+r_{sp1}r_{ps2}\\
D_{2} & =r_{ss1}r_{sp2}+r_{sp1}r_{pp2}\\
D_{3} & =r_{ps1}r_{ss2}+r_{pp1}r_{ps2}\\
D_{4} & =r_{ps1}r_{sp2}+r_{pp1}r_{pp2}
\end{align*}
with these definitions of $D_1$, $D_2$, $D_3$ and $D_4$, 
\begin{equation}
L=1-(D_{1}+D_{4})e^{-2k_{1}d}+(D_{1}D_{4}-D_{2}D_{3})e^{-4k_{1}d}
\end{equation}
The derivative of $\mathrm{Ln}(L)$, $-\frac{\partial \mathrm{Ln}(L)}{\partial d}$, is given by
\begin{equation}
\frac{-2k_{1}}{L}[(D_{1}+D_{4})e^{-2k_{1}d}-2(D_{1}D_{4}-D_{2}D_{3})e^{-4k_{1}d}]  \label{deriv}
\end{equation}
For a non-gyrotropic case $r_{sp1}=r_{ps1}=0$ and $r_{sp2}=r_{ps2}=0$,
so we have $D_{1}=r_{ss1}r_{ss2}$, $D_{4}=r_{pp1}r_{pp2}$, $D_{2}=D_{3}=0$ and $L$ is
\begin{equation}
1-(r_{ss1}r_{ss2}+r_{pp1}r_{pp2})e^{-2k_{1}d}+r_{ss1}r_{ss2}r_{pp1}r_{pp2}e^{-4k_{1}d}
\end{equation}
The derivative of $\mathrm{Ln}(L)$ becomes, 
\begin{equation}
-\frac{\partial \mathrm{Ln}(L)}{\partial d}=\frac{-2k_{1}}{L}[(D_{1}+D_{4})e^{-2k_{1}d}-2(D_{1}D_{4})e^{-4k_{1}d}]
\end{equation}
Then a repulsive force $F>0$ requires the following, 
\begin{equation}
\sum_{n}\int\frac{d^{2}k}{(2\pi)^{2}} \frac{2k_{1}}{L}e^{-2k_{1}d}[(D_{1}+D_{4})-2D_{1}D_{4}e^{-2k_{1}d}]<0
\end{equation}
Known that $k_{1}>0$, $L>0$ and $e^{-2k_{1}d}>0$, we use an approximation that $\frac{2k_{1}}{L}e^{-2k_{1}d}$ does not change too much in the sum of $n$ and integral over $k$.
The qualitative condition for a repulsive force is
\begin{equation}
\sum_{n}\int\frac{d^{2}k}{(2\pi)^{2}}[(D_{1}+D_{4})-2D_{1}D_{4}e^{-2k_{1}d}]<0
\end{equation}
If the absolute value of the reflection coefficient satisfy the condition
$|r_{ss}|<1$ and $|r_{pp}|<1$, known that $e^{-2k_{1}d}<1$, we
have $|r_{ss1}r_{ss2}|>|r_{ss1}r_{ss2}r_{pp1}r_{pp2}|e^{-2k_{1}d}$, and $|r_{pp1}r_{pp2}|>|r_{ss1}r_{ss2}r_{pp1}r_{pp2}|e^{-2k_{1}d}$, so $|D_{1}|+|D_{4}|>2|D_{1}D_{4}|e^{-2k_{1}d}$, the dominat term is $D_{1}+D_{4}$, the repulsive force condition simplifies as
\begin{equation}
\sum_{n}\int\frac{d^{2}k}{(2\pi)^{2}}(r_{ss1}r_{ss2}+r_{pp1}r_{pp2})<0
\end{equation}
For an ultra-strong gyrotropic material with giant polarization interconversion, we need to start from the Eqn.~(\ref{deriv}). In this case, the reflection
coefficients $|r_{sp}|$ and $|r_{ps}|$ could be much larger than
$|r_{ss}|$ and $|r_{pp}|$, $D_{1}\simeq r_{sp1}r_{ps2}$, $D_{4}\simeq r_{ps1}r_{sp2}$
and $D_{2}\simeq0$, $D_{3}\simeq0$, follow the same logic we have
\begin{equation}
\sum_{n}\int\frac{d^{2}k}{(2\pi)^{2}}(r_{sp1}r_{ps2}+r_{ps1}r_{sp2})<0
\end{equation}

\section{Casimir force between a perfectly conducting plate and an infinitely permeable plate }
For a perfectly conducting plate $\epsilon=\infty$ and $\mu=1$, and for an infinitely permeable plate $\mu=\infty$ and $\epsilon=1$.
The reflection coefficients at the interface of a perfectly conducting plate and a vacuum is $r_{ss}=-1$ and $r_{pp}=1$.
The reflection coefficients at the interface of an infinitely permeable plate and a vacuum is $r_{ss}=1$ and $r_{pp}=-1$. For both cases $r_{sp}=r_{ps}=0$, so no polarization interconversion happens.
At zero temperature, the sum of Matsubara freqencies becomes an integral $2\pi k_{B}T/\hbar\times\sum_{n}=\int d\xi$, assume the system is isotropic, so the integral over the angle $\theta$ gives $2\pi$, the Casimir pressure is
\begin{equation}
P=\hbar/(2\pi)\int d\xi\int\frac{2\pi kdk}{(2\pi)^{2}}\left[-\frac{\partial \mathrm{Ln}(L)}{\partial d}\right]
\end{equation}
We consider two cases below:

i) Casimir force between two perfectly conducting plates, 
\begin{equation}
L=1-2e^{-2k_{1}d}+e^{-4k_{1}d}
\end{equation}
The derivative of $\mathrm{Ln}(L)$ becomes, 
\begin{equation}
-\frac{\partial \mathrm{Ln}(L)}{\partial d}=\frac{-2k_{1}}{L}[2e^{-2k_{1}d}-2e^{-4k_{1}d}]
\end{equation}
The attractive Casimir pressure is
\begin{equation}
P_C=-\frac{\hbar}{\pi^{2}}\int d\xi\int k_{1}kdk\frac{e^{-2k_{1}d}}{1-e^{-2k_{1}d}}
\end{equation}
Known that $k_{1}^2=k^2+(\xi/c)^2$, we could introduce polar coordinates according to $k=k_1 \mathrm{sin}(\phi)$ and  $\xi/c=k_1 \mathrm{cos}(\phi)$, so $\int^{\infty}_0 d(\xi/c) \int^{\infty}_0 dk=\int^{\infty}_0 k_1dk_1 \int^{\pi/2}_0 d\phi$, then the Casimir pressure is
\begin{equation}
P_C=-\frac{\hbar c}{\pi^{2}}\int^{\infty}_0 k_1^3 dk_1 \int^{\pi/2}_0 \mathrm{sin}(\phi) d\phi \frac{e^{-2k_{1}d}}{1-e^{-2k_{1}d}}
\end{equation}
Note that the integral
\begin{equation}
\int^{\infty}_0 dx\frac{x^3 e^{-2x}}{1-e^{-2x}}=\frac{\pi^4}{240}
\end{equation}
we have
\begin{equation}
P_C=-\frac{\hbar c \pi^{2}}{240 d^4}
\end{equation}
ii) Casimir force between a perfectly conducting plate and an infinitely permeable plate, 
\begin{equation}
L=1+2e^{-2k_{1}d}+e^{-4k_{1}d}
\end{equation}
The derivative of $\mathrm{Ln}(L)$ becomes, 
\begin{equation}
-\frac{\partial \mathrm{Ln}(L)}{\partial d}=\frac{-2k_{1}}{L}[-2e^{-2k_{1}d}-2e^{-4k_{1}d}]
\end{equation}
The repulsive Casimir pressure is
\begin{equation}
P_B=\frac{\hbar}{\pi^{2}}\int d\xi\int k_{1}kdk\frac{e^{-2k_{1}d}}{1+e^{-2k_{1}d}}
\end{equation}
Follow the same logic in case i) and note that the integral
\begin{equation}
\int^{\infty}_0 dx\frac{x^3 e^{-2x}}{1+e^{-2x}}=\frac{7\pi^4}{1920}
\end{equation}
we have
\begin{equation}
P_B=\frac{7 \hbar c \pi^{2}}{1920 d^4}=-(7/8)P_C
\end{equation}
For a typical distance $d=0.2\mu m$, $P_C=-0.8$ Pa.

\section{Reflection matrix between a vacuum and a gyrotropic plate}
For more general cases with polarization interconversion, we solve the reflection coefficients in the way below. Assume the
amplitudes of TM mode and TE mode are $A_{TM}$ and $A_{TE}$ respectively, the injecting electromagnetic wave in the vacuum is, \begin{widetext}
\begin{eqnarray*}
E_{0} & = & (k_{z}A_{TM}/\omega,A_{TE},-k_{x}A_{TM}/\omega)e^{ik_{x}x+ik_{z}z-i\omega t}\text{ for }z>0
\end{eqnarray*}
\begin{eqnarray*}
H_{0} & = & (-k_{z}A_{TE}/\omega,A_{TM},k_{x}A_{TE}/\omega)e^{ik_{x}x+ik_{z}z-i\omega t}\text{ for }z>0
\end{eqnarray*}
The reflected wave is 
\begin{eqnarray*}
E_{r} & = & (-k_{z}R_{TM}/\omega,R_{TE},-k_{x}R_{TM}/\omega)e^{ik_{x}x-ik_{z}z-i\omega t}\text{ for }z>0
\end{eqnarray*}
\begin{eqnarray*}
H_{r} & = & (k_{z}R_{TE}/\omega,R_{TM},k_{x}R_{TE}/\omega)e^{ik_{x}x-ik_{z}z-i\omega t}\text{ for }z>0
\end{eqnarray*}
The transmitted wave inside a specific material is given by, 
\begin{eqnarray*}
E_{t} & = & (e_{x},e_{y},e_{z})e^{ik_{x}x+iqz-i\omega t}\text{ for }z<0
\end{eqnarray*}
\begin{eqnarray*}
H_{t} & = & (h_{x},h_{y},h_{z})e^{ik_{x}x+iqz-i\omega t}\text{ for }z<0
\end{eqnarray*}
To obtain the Fresnel coefficients for a gyrotropic slab, we first
consider the injecting TE mode to be zero ($A_{TE}=0$) and write
down the boundary conditions 
\begin{eqnarray}
(k_{z}/\omega)(A_{TM}-R_{TM})=A_{1}e_{x1}+A_{2}e_{x2}\nonumber \\
R_{TE}=A_{1}e_{y1}+A_{2}e_{y2}\nonumber \\
(k_{z}/\omega)(R_{TE})=A_{1}h_{x1}+A_{2}h_{x2}\nonumber \\
(A_{TM}+R_{TM})=A_{1}h_{y1}+A_{2}h_{y2}
\end{eqnarray}
from which we obtain the ratio of $A_{1}$ and $A_{2}$, 
\begin{equation}
[(k_{z}/\omega)e_{y1}-h_{x1}]A_{1}=A_{2}[h_{x2}-(k_{z}/\omega)e_{y2}]
\end{equation}
where $\psi_{1}=[e_{x1},e_{y1},e_{z1},h_{x1},h_{y1},h_{z1}]^{T}=[E_{1},H_{1}]^{T}$
and $\psi_{2}=[e_{x2},e_{y2},e_{z2},h_{x2},h_{y2},h_{z2}]^{T}=[E_{2},H_{2}]^{T}$
are the transmitted electromagnetic wave function inside a specific
medium (gyrotropic or magneto-electric), usually obtained numerically.
We also obtain the following equations, 
\begin{eqnarray}
(k_{z}/\omega)(A_{TM}+R_{TM})=A_{1}(k_{z}/\omega)h_{y1}+A_{2}(k_{z}/\omega)h_{y2}\nonumber \\
2(k_{z}/\omega)(A_{TM})=A_{1}[(k_{z}/\omega)h_{y1}+e_{x1}]+A_{2}[(k_{z}/\omega)h_{y2}+e_{x2}]\nonumber \\
2(k_{z}/\omega)(R_{TM})=A_{1}[(k_{z}/\omega)h_{y1}-e_{x1}]+A_{2}[(k_{z}/\omega)h_{y2}-e_{x2}]
\end{eqnarray}
the Fresnel coefficients $r_{pp}$ and $r_{ps}$ are obtained, 
\begin{eqnarray}
r_{pp}=R_{TM}/A_{TM}=\frac{A_{1}[(k_{z}/\omega)h_{y1}-e_{x1}]+A_{2}[(k_{z}/\omega)h_{y2}-e_{x2}]}{A_{1}[(k_{z}/\omega)h_{y1}+e_{x1}]+A_{2}[(k_{z}/\omega)h_{y2}+e_{x2}]}\\
r_{ps}=R_{TE}/A_{TM}=\frac{2(A_{1}h_{x1}+A_{2}h_{x2})}{A_{1}[(k_{z}/\omega)h_{y1}+e_{x1}]+A_{2}[(k_{z}/\omega)h_{y2}+e_{x2}]}
\end{eqnarray}
Then we consider the injecting TM mode to be zero ($A_{TM}=0$), the
boundary conditions are given by, 
\begin{eqnarray}
(k_{z}/\omega)(-R_{TM})=C_{1}e_{x1}+C_{2}e_{x2}\nonumber \\
(A_{TE}+R_{TE})=C_{1}e_{y1}+C_{2}e_{y2}\nonumber \\
(k_{z}/\omega)(R_{TE}-A_{TE})=C_{1}h_{x1}+C_{2}h_{x2}\nonumber \\
(R_{TM})=C_{1}h_{y1}+C_{2}h_{y2}
\end{eqnarray}
from which we obtain the ratio of $C_{1}$ and $C_{2}$, 
\begin{equation}
[(k_{z}/\omega)h_{y1}+e_{x1}]C_{1}=-C_{2}[(k_{z}/\omega)h_{y2}+e_{x2}]
\end{equation}
and the following equations, 
\begin{eqnarray}
(k_{z}/\omega)(A_{TE}+R_{TE})=C_{1}(k_{z}/\omega)e_{y1}+C_{2}(k_{z}/\omega)e_{y2}\nonumber \\
2(k_{z}/\omega)(A_{TE})=C_{1}[(k_{z}/\omega)e_{y1}-h_{x1}]+C_{2}[(k_{z}/\omega)e_{y2}-h_{x2}]\nonumber \\
2(k_{z}/\omega)(R_{TE})=C_{1}[(k_{z}/\omega)e_{y1}+h_{x1}]+C_{2}[(k_{z}/\omega)e_{y2}+h_{x2}]
\end{eqnarray}
the Fresnel coefficients $r_{ss}$ and $r_{sp}$ are obtained, 
\begin{eqnarray}
r_{ss}=R_{TE}/A_{TE}=\frac{C_{1}[(k_{z}/\omega)e_{y1}+h_{x1}]+C_{2}[(k_{z}/\omega)e_{y2}+h_{x2}]}{C_{1}[(k_{z}/\omega)e_{y1}-h_{x1}]+C_{2}[(k_{z}/\omega)e_{y2}-h_{x2}]}\\
r_{sp}=R_{TM}/A_{TE}=\frac{-2(C_{1}e_{x1}+C_{2}e_{x2})}{C_{1}[(k_{z}/\omega)e_{y1}-h_{x1}]+C_{2}[(k_{z}/\omega)e_{y2}-h_{x2}]}
\end{eqnarray}

\end{widetext}

\end{document}